# The Coronal Mass Ejection Visibility Function of Modern Coronagraphs

Angelos Vourlidas[1,2], L. A. Balmaceda[3,6], H. Xie[4,6], O. C. St. Cyr[5,6]

[1] The Johns Hopkins University Applied Physics Laboratory, Laurel, MD 20723, USA
[2] IAASARS, Observatory of Athens, Penteli, Greece
[3] George Mason University, Fairfax, VA, USA
[4] Catholic University of America, Washington D.C.
[5] Universities Space Research Association, Columbia, MD 21046, USA
[6] NASA, Goddard Space Flight Center, Heliophysics Science Division, Greenbelt, MD 20771, USA

## ABSTRACT
We analyze the detection capability of Coronal Mass Ejections (CMEs) for all currently operating coronagraphs in space. We define as CMEs events that propagate beyond 10 solar radii with morphologies broadly consistent with a magnetic flux rope presence. We take advantage of multi-viewpoint observations over five month-long intervals, corresponding to special orbital configurations of the coronagraphs aboard the STEREO and SOHO missions. This allows us to sort out CMEs from other outward-propagating features (e.g. waves or outflows), and thus to identify the total number of unique CMEs ejected during those periods. We determine the CME visibility functions of the STEREO COR2-A/B and LASCO C2/C3 coronagraphs directly as the ratio of observed to unique CMEs. The visibility functions range from 0.71 to 0.92 for a 95% confidence interval. By comparing detections between coronagraphs on the same spacecraft and from multiple spacecraft, we assess the influence of field of view, instrument performance, and projection effects on the CME detection ability without resorting to proxies, such as flares or radio bursts. We find that no major CMEs are missed by any of the coronagraphs, that a few slow halo-like events may be missed in synoptic cadence movies and, that narrow field of view coronagraphs have difficulties discriminating between CMEs and other ejections leading to 'false' detection rates. We conclude that CME detection can only be validated with multi-viewpoint imaging-- two coronagraphs in quadrature offer adequate detection capability. Finally, we apply the visibility functions to observed CME rates resulting in upward corrections of 40%.

## 1. Introduction

Coronal Mass Ejections (CMEs) are large scale expulsions of magnetized plasma resulting from the explosive release of magnetic energy stored in the solar atmosphere. They are also the primary drivers of terrestrial and planetary space weather. As CMEs propagate at speeds upwards of several hundred to a few thousand km/s, they interact with planetary bodies and spacecraft throughout the heliosphere. It is, therefore, important to have a good understanding of CME properties (i.e. rates of occurrence, kinematic profiles, energy content, etc.) to better quantify solar activity to improve forecasting of space weather.

Because coronagraphs detect CMEs via the scattering of photospheric visible light from the electrons in the CME plasma, the visibility of a given CME depends strongly on its direction of propagation with respect to an observer. The scattering efficiency falls off as the square of the cosine of the angular distance from the plane-of-sky (POS) and hence events propagating away from the POS tend to be fainter than events propagating along the POS (e.g. Figure 2, Morrill et al. 2009). The effect is stronger for polarization brightness (pB) compared to total brightness observations (Vourlidas & Howard 2006) and is of concern primarily for ground-based coronagraphic observations, which are overwhelming performed in pB, soon to be joined in space by the Polarimeter to Unify the Corona and Heliosphere (PUNCH; DeForest et al. 2019). Therefore, the detection of a CME by a given coronagraph depends both on the event-observer geometry and on the instrumental sensitivity and stray-light level of the detecting coronagraph. This dependence is called the *CME Visibility Function (VF)*, and is specific to a given coronagraph.

The visibility function is used primarily to correct CME occurrence rates, which can then be compared to other solar features, such as flares or sunspots, and help understand the connections among the various manifestations of solar activity (Webb & Howard 1994; St. Cyr et al. 2015; Webb et al. 2017). Since the CME mass can be derived from the observations, improved VFs can also lead to better estimates of the CME contribution to the total mass loss from the Sun and help constrain the mass limits, and by extension, kinetic energies of CMEs.

The highly sensitive CCD detectors in the Large Angle and Spectroscopic Coronagraph (LASCO; Brueckner et al. 1995), along with the 24x7 observations of the corona revolutionized CME studies. The coronagraphs detected faint halo CMEs from the beginning of their science operations, leading to the adoption of the coronagraph as a key observational capability for Space Weather.

It is, however, difficult to derive a VF for a coronagraph since there is no independent method of detecting CMEs to know their total number. Researchers in the past relied on other forms of solar activity, such as flares (Yashiro et al. 2005), radio type II emission (Webb & Howard 1994) or EUV post-eruptive arcades (Tripathi et al. 2004) as proxies for CMEs, Unfortunately, there is no one-to-one correspondence between a CME and any of these proxies. There exist CMEs without obvious low corona manifestations, like post-eruptive arcades (so called 'stealth-CMEs'; Robbrecht et al. 2009) and large flares can occur without a CME (e.g. Thalmann et al. 2015). In addition, behind-the-limb CMEs are detected by coronagraphs but their low-coronal sources are occulted. Because of the lack of independent validation, the association rates between CMEs and a proxy, say, flares are uncertain. For example, Webb & Howard (1994) assumed that all CMEs associated with metric type II from regions close to the limb (60º- 90º from central meridian) can be detected. They then used that fraction (62%) to extrapolate to the remaining longitudes and hence derive VFs and correct occurrence rates for *Skylab*, *Solwind* and *SMM*. In the first analysis of CME detectability from the LASCO coronagraphs, St Cyr et al (2000) undertook a preliminary VF analysis study based on Type-IIs and claimed: "This indicates that little, if any, correction will be required as a visibility function for detection of CMEs for LASCO". Yashiro et al. (2005) had to assume that all CMEs associated with limb flares were detected in order to estimate a CME-flare association rate and a VF for LASCO. They concluded that 25%-67% of CMEs associated with C-class flares were invisible but there is no way to validate this result without an independent detection of these CMEs. None of these studies, however, can be considered as true instrument VFs.

Since early 2007, the Sun-Earth Connection Coronal and Heliospheric Investigation (SECCHI; Howard et al. 2008) coronagraphs aboard the *Solar Terrestrial Relations Observatory* (STEREO; Kaiser et al. 2008) mission, along with the LASCO coronagraphs, provide near-simultaneous CME observations from three vantage points (see Sec. 2 for details). For the first time since the discovery of CMEs, we are in the position to have independent measurements of the same CME from a different instrument and different viewpoint (e.g. Vourlidas et al. 2017; Balmaceda et al. 2018; Kwon et al. 2015). Hence, we should now be able to derive the true VF for the STEREO and LASCO coronagraphs.

Recently, Bronarska et al. (2017) attempted to do that for the LASCO coronagraphs using existing event lists (a manual list for LASCO and an automated list for SECCHI) for June-November 2011. Although it is a reasonable approach for a quick assessment of the LASCO VF, it cannot be considered a robust result. The reliance on event lists (and the associated parameters such as position angle, widths, time), instead of visually identifying an event in all three coronagraphs, has serious shortcomings. The CDAW list, for example, contains as much as 30% of questionable entries (e.g. wrong identifications, poor visibility events or multiple entries for the same event---see Vourlidas et al. 2013) while the CACTus SECCHI CME list is heavily biased on the threshold selection criteria for the algorithm (Hess & Colaninno 2017). Many of the most questionable events may not even be CMEs, in the sense that they may be magnetically driven eruptions of solar plasma, as expected from current theories. For this reason, a precise definition of a CME is needed. We discuss this definition in detail in Section 2.2. In addition, the number of unique events---a critical quantity for the proper VF calculation---cannot be derived from such lists. A visual assessment is required. Therefore, until this work, there have been no robust measurements of VFs available.

The work reported here aims to provide a thorough analysis of CME detectability for all currently operating space-based coronagraphs, over different phases of the solar cycle, and to interpret the VF results in terms of instrument design, performance, and effects of projection. Part of the motivation behind the determination of a VF is the reconciliation of CME rates among coronagraphs over multiple solar cycles (Webb et al. 2017) and the analysis presented here is a step towards that goal.

The paper is organized as follows. In the next section, we explain our data sources, selection criteria for the study periods, our methodology, and VF definition. In Section 3, we present the results as a function of viewing geometry and field of view (FOV). We discuss the implications of the results in Section 4 for the Visibility Function, instrument design, and space weather. We conclude in Section 5.

## 2. Event Selection and Methodology

The CME observations are from the SECCHI/COR1 and COR2 coronagraphs aboard STEREO-A and –B (STA and STB, hereafter), and the LASCO/C2 and C3 coronagraphs aboard the SOHO mission over the 2007-2011 period. The nominal COR1 and COR2 fields of view (FOV) extend from 1.5 to 4 Rs and 2.5 to 15 Rs, respectively. The LASCO/C2 and C3 FOVs are 2.2-6

Rs, and 3.7-32 Rs, respectively. We use only total brightness images taken at the synoptic cadence for each telescope; namely, at 5 min, 15 min, and 12 mins, for COR1, COR2 and C2/C3, respectively. The images are 'quicklook' images, widely used across the community for event identification and kinematic analysis. 'Quicklook' images are the ratio of a given image to a monthly average. The operation removes the F-corona and stray-light contribution and corrects for instrument vignetting and thus corrects for intensity drop-off with distance. We rely on running and base difference, and direct images, as appropriate, to identify the CMEs and their morphology.

## 2.1 Event Selection

Because of the large amount of work involved in visually identifying and cataloguing CMEs in multiple instruments, we focus our analysis on one-month periods when special viewing configurations between the SECCHI and LASCO coronagraphs arise. We wanted durations that were at least a solar rotation in length, and we believe this provides a sufficient snap-shot at different times of the solar cycle. We selected these specific time frames when data gaps were minimal and the observing duty cycles were fairly complete. The periods are:

(1) **April-May 2007**. The SECCHI and LASCO are nearly aligned along the Sun-Earth line. Because of the near-identical viewing geometry (the maximum separation between STA and STB is only 7°), this period allows us to cross-calibrate the detection levels between LASCO and SECCHI. We analyze two months to build a large enough sample due to the low solar activity.

(2) **December 2007**. The SECCHI coronagraphs are about 45° apart and are ~20° from LASCO. Because the Thomson scattering efficiency of the electrons entrained in the CME drops sharply beyond around 40° (see Figure 3 in Vourlidas et al. 2010), we expect to witness this effect in the CME detection rate and morphology between SECCHI-A and SECCHI-B but not between LASCO and either of the SECCHI coronagraphs.

(3) **April 2009,** STA and STB are in quadrature and are ~45° from LASCO. Projection effects (due to Thomson scattering) maximize at quadrature. For example, a halo-like CME for STA will be a limb CME for STB, and vice versa. In addition, STA and STB are approaching the Sun-Earth $L_4$ and $L_5$ Lagrangian points and we could extend this work to evaluate the benefit of observations from those locations for Space Weather research (see Vourlidas (2015) and references therein).

(4) **May 2011**. SECCHI and LASCO are in quadrature while STA-STB are in opposition. This configuration is similar to the April 2009 period but now STA and STB have the same viewing geometry allowing cross-check. Thompson et al. (2011) used the opposition configuration for inter-calibration of COR1-A/B and showed they were mirror images. Hence, COR1 (and COR2) should lose the dual-viewpoint advantage, reducing to an effectively single instrument.

There is another interesting configuration when the three spacecraft are about 120° apart (around August-September 2012). In this geometry, a CME of any width or direction should be detectable by one of the telescopes, assuming it is above their detectability threshold and fast enough to be detected with standard cadence programs (see Section 4.3 discussion). However, the analysis of that period does not add anything fundamentally new to this work, as we discuss later. We choose to keep the paper focused on the four periods discussed above.

## 2.2 CME Definition

To properly understand our methodology and results, it is important to explain what we consider a CME in this work. The CME definition is still debated despite 50 years of CME observations. The original definition given by Hundhausen et al (1984) as modified by Schwenn (2006) is: *"We define a coronal mass ejection to be an observable change in coronal structure that (1) occurs on a time scale of a few minutes and several hours and (2) involves the appearance (and outward motion) of a new, discrete, bright, white light feature in the coronagraph field of view."*

Vourlidas et al. (2013) argued that this definition, well-suited for the 1970's when CMEs were a novel phenomenon, is no longer satisfactory. We know a lot more about CMEs with rather firm theoretical and modeling expectations. Every model/theory expects a CME to be the result of the ejection of a magnetic flux rope (MFR) (see reviews by Chen 2011 and Liu 2020). Many comparisons between coronagraph images and modeling over the years have established the expected signatures of MFRs in the images (see examples in Vourlidas et al. 2013; Vourlidas 2014).

As Schwenn (2006) updated the original CME definition by adding '(and outward motion)', so did Vourlidas et al. (2013) by incorporating the theoretical developments and statistical results from the observations of thousands of events in the following definition: *"We define a 'Flux Rope'-CME to be the eruption of a coherent magnetic, twist-carrying coronal structure with angular width of at least 40° and able to reach beyond 10 Rs, which occurs on a time scale of a few minutes to several hours."*

We start with this definition here but we have to relax it for practical reasons. We deal with single viewpoint observations hence we do not know the true width of the events. Therefore, we have to consider all widths. We then take into account the morphological classification of each event (discussed at length in 2.4 and also in Vourlidas et al. 2013, 2017) to assess whether the event may contain a MFR. Essentially, we consider as a CME any event with clear or likely flux rope structure that propagates beyond 10 Rs. By adopting this more nuanced CME definition we focus our analysis to events that are most important in basic research and space weather operations and hence make our conclusions relevant and useful to the widest possible audience.

## 2.3 Visibility Function Concerns

We define the Visibility Function for a given coronagraph as the ratio of detected to total number of CMEs over a given period (about a month, in our case). This ratio depends primarily on the sensitivity of the instrument or signal-to-noise ratio (SNR) of the synoptic images. The SNR, in turn, depends on the stray light level, exposure time, detector efficiency, and overall system throughput. Of course, the SNR also depends on the excess brightness of the CME over the K-corona, F-corona, and instrumental backgrounds. Since the CME brightness is a function of both the amount of plasma within the transient and the CME-observer geometry, the latter needs to be considered in the VF analysis.

The biggest source of uncertainty is the unknown total number of CMEs because a CME event is traditionally defined by its detection by a white light coronagraph, which is the instrument we want to estimate the VF of. There is no independent way to know that an event has occurred and is propagating in the corona without observations from another coronagraph and viewpoint.

Obviously, we are now in position to resolve this conundrum thanks to the SECCHI instruments. But the COR1/COR2 coronagraphs have different FOVs than LASCO and we need to consider the FOV effect, both between COR1 and COR2 (and LASCO C2 and C3) and between SECCHI and LASCO. Therefore, we break our VF analysis in two areas: geometric effects and FOV effects. For the geometric effects, we compare COR2-A, COR2-B, and LASCO (treated as a single instrument here). For the FOV effects, we consider COR1-COR2 for STA and STB separately.

## 2.4 Methodology

We use the existing online CME catalogs for LASCO (CDAW[1]) and SECCHI (COR1[2] and MVC[3]) only as a reference as they are not strictly appropriate for this work as discussed above. To properly evaluate the FOV, instrumental, and projection effects, the observations from each coronagraph are inspected separately and new CME lists are constructed for each telescope (links to the lists are given in Table 1). Then, we proceed to make two preliminary joint lists (COR1-COR2) and (COR2-A-LASCO-COR2-B) for each of the four time periods. We consider LASCO as a single coronagraph for the visibility function work but we do compare events in C2 and C3, as we discuss in 3.2.2.

We then inspect movies in all coronagraphs to understand why any events were missed. We assess the morphology based on previous classifications (Vourlidas et al. 2013, 2017). Although these are subjective classifications, they indirectly reflect the physical nature of the event as we have argued in those publications. We assign a letter code for each missed event to identify the reason. The letter codes are: G for data gap, NR for narrow event, FL for failed, OVR for the case of overlapping events, S for slow (< 200 km/s) events, and FN for faint events. The other codes in the online lists are described in Vourlidas et al. (2017) and the MVC webpage. The rationale for these codes is as follows.

Narrow events (i.e. jets) directed towards a given coronagraph may be occulted for long distances and/or may be too faint to be detected. 'Failed' refers to a class of events discussed by Vourlidas et al. (2010) and identified originally by St. Cyr et al. (2000), where events bright enough to be detected in the low corona, dim rapidly and disappear within the COR2 or C3 FOV. We have attributed these events to wave fronts (Vourlidas et al. 2017) and hence we do not consider them as CMEs. These are the only types of events that we discard from the analysis because they are the least likely to carry an MFR.

A CME may be missed by a given coronagraph if the event overlaps with other event(s) in progress in that coronagraph's FOV. This is a relatively common situation during solar maximum. Slow events, such as Streamer-Blowouts (Vourlidas & Webb 2018), if directed towards a coronagraph, can be missed using the routine running difference detection schemes based on synoptic cadences. Although these events are not 'missed', in a strict sense, they would not be detected by the casual observer, a forecaster or anyone else without some prior knowledge that an event has occurred. Such events are frequently associated with 'stealth' CMEs (Robbrecht et al. 2009) and lack the usual coronal and lower atmosphere signatures that accompany eruptions. Generally speaking, they can only be detected with multi-viewpoint coronagraphic observations.

---

[1] https://cdaw.gsfc.nasa.gov/CME_list/
[2] http://cor1.gsfc.nasa.gov/catalog/
[3] http://solar.jhuapl.edu/Data-Products/COR-CME-Catalog.php

We use the term 'faint' in a visual rather than a strictly quantitative sense. We refer to events that are difficult to discern in running difference movies, even after considerable experimentation with the display levels. We make an effort to assess whether the events are inherently faint (i.e. less dense) or faint due to projection effects (i.e. events propagating away from the sky plane). We reserve the term for the former and use other terms such as narrow or overlapping to mark detections affected by projection effects. The rationale for data gaps is obvious.

The identification and assignments are done independently by two of the authors (L.B and H.X) and the whole team convenes to discuss discrepancies and agree on the final list and classifications. We assume that the number of unique events (whether they were seen by one or more of the three telescopes) is the absolute total number of CMEs during that period. We discuss the validity of this assumption in Section 4.

## 3. Results

In this section, we present the results of CME detections across spacecraft ('Viewpoint Effects') and across instruments on the same spacecraft ('FOV Effects') along with a discussion of the types of CMEs missed for each case.

### 3.1. Viewpoint Effects

Table 1 CME Detections and Visibility Functions of the LASCO and SECCHI Coronagraphs. The visibility function, given by the ratio of detected to unique events, is given in boldface in the parentheses. nG denotes the number of events, n, occurring during a telescope's data gap. We assume that the events during data gaps would have been detected. The links to the event lists are given at the end of the table.

| Date | Unique Events | COR2-A | COR2-B | LASCO |
|---|---|---|---|---|
| Apr 2007[1] | 16 | 16 **(1)** | 12 **(0.75)** | 16 **(1)** |
| May 2007[2] | 35 | 35 **(1)** | 29 **(0.83)** | 32 + 3G **(1)** |
| Dec 2007[3] | 22 | 20 **(0.91)** | 12+ 3G **(0.68)** | 22 **(1)** |
| Apr 2009[4] | 22 | 19+ 1G **(0.91)** | 14+ 2G **(0.73)** | 19+ 1G **(0.91)** |
| May 2011[1] | 91 | 74+ 4G **(0.86)** | 73 **(0.80)** | 75 **(0.82)** |

[1] https://cor1.gsfc.nasa.gov/catalog/corc2/com_list_200704_corc2_final.html
[2] https://cor1.gsfc.nasa.gov/catalog/corc2/com_list_200705_corc2_final_rev.html
[3] https://cor1.gsfc.nasa.gov/catalog/corc2/com_list_200712_corc2_final_rev.html
[4] https://cor1.gsfc.nasa.gov/catalog/corc2/com_list_200712_corc2_final_rev.html
[5] https://cor1.gsfc.nasa.gov/catalog/corc2/com_list_201105_corc2_final.html

The results for COR2 and LASCO are summarized in Tables 1 and 2. COR2-A and LASCO have very similar VFs, ranging from 0.82 to 1, while the COR2-B VF is considerably lower. This is expected as COR2-B has a higher level of instrumental stray light than COR2-A. The effect has been discussed in a bit more detail in Frazin et al. (2012) and Vourlidas et al. (2017). We expand the VF discussion and its implications for CME rates in 4.1.

Turning to the nature of the missed events (Table 2), we find that the majority of them consist of faint narrow events (14), followed by failed (7) and faint slow events (5). Naturally, most of these faint events occur during solar minimum (2007-09) when the average CME mass (and hence brightness) is reduced. Since we do not consider 'failed' events as CMEs, we will not include them in the following discussion. The total number of missed events is 68 for all three telescopes.

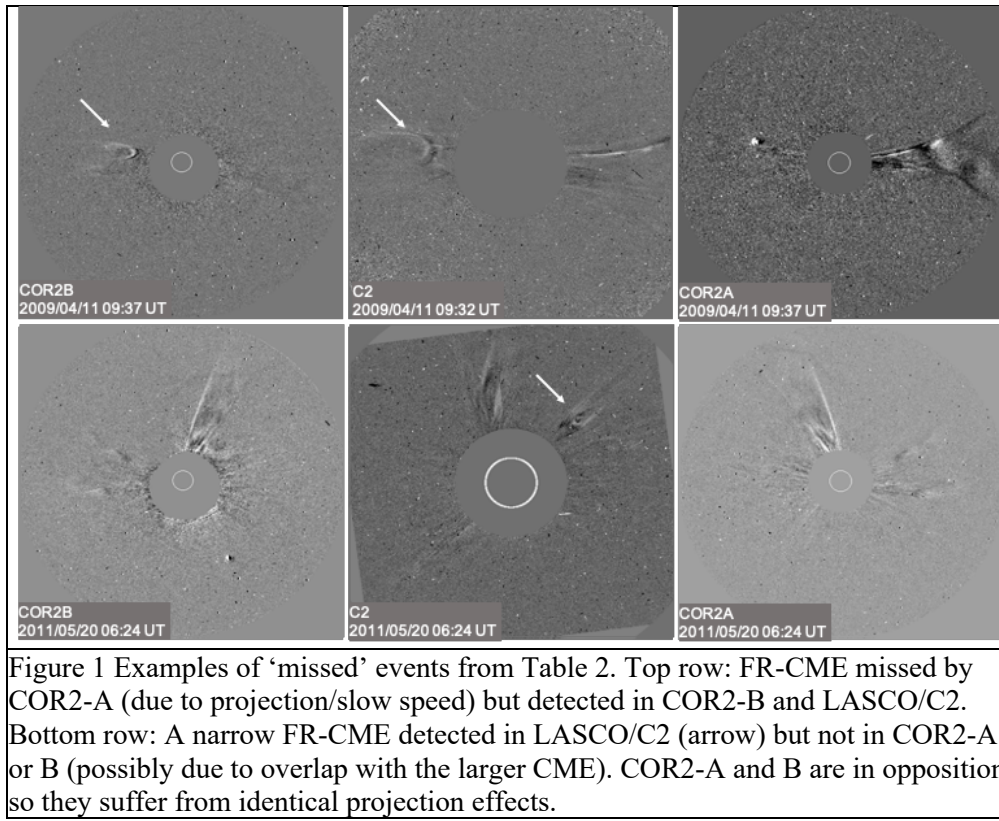

Figure 1 Examples of 'missed' events from Table 2. Top row: FR-CME missed by COR2-A (due to projection/slow speed) but detected in COR2-B and LASCO/C2. Bottom row: A narrow FR-CME detected in LASCO/C2 (arrow) but not in COR2-A or B (possibly due to overlap with the larger CME). COR2-A and B are in opposition so they suffer from identical projection effects.

As an example of the types of CMEs that are listed as 'missed' in Table 2, we present, in Figure 1, two such examples. The top row shows a CME (flux-rope type per Vourlidas et al. 2017) seen over the east limbs of COR2B (left) and C2 (middle). It is not seen in COR2A (right) because it is a halo from that viewpoint and it is too faint and slow to be detected in standard cadence running difference images. We attribute this missed event to projection effects. In the bottom row is an example from May 20, 2011 when COR2A and B are in opposition. All three coronagraphs detect a CME along the solar north but only C2 detects the CME, marked by the arrow, as a narrow, flux-rope type ejection along the northwest. We attribute this missed event to its narrow width and possible overlap with the larger CME along the COR2A/B lines of sight. Their opposition effectively reduces the two COR2s to a single viewpoint.

The important, if not dominant, role of observing geometry is clearly revealed in the quadrature comparison (May 2011) and to a lesser degree in April 2009. We find that almost all events (46 out of 56, in 2009-2011) missed by a coronagraph are seen by the other coronagraph in quadrature. STA and STB, being in opposition, effectively reduce to a single spacecraft as far as CME detection is concerned. This result essentially confirms our hypothesis (Sec. 2.1) that quadrature observations are a powerful diagnostic of projection effects in CME detection. Cremades et al. (2015) reached a similar conclusion from a center-to-limb study of CMEs from a single active region complex.

Examining the results in Table 2 from a wider perspective provides further insights. Not all of the events marked as faint were missed by the telescope with the lowest SNR (COR2-B). So, we cannot attribute all of them to low instrumental performance. Particularly, the faint narrow events (NR/FN) suggest that the viewing geometry is also to blame so their detectability should be included under the projection effects bin. Adding to this sum, the overlapped event (since it was clearly detected from another viewpoint), brings the ratio of events missed due to some sort of projection effect to 84% (57/68). We conclude, therefore, that the brightness of an event *as viewed from a given vantage point* is the primary consideration for its detectability.

Table 2 Characteristics of the CME events missed by a given telescope.

| | | Missed Events | CME Characteristics* | Projection Effect |
|---|---|---|---|---|
| April 2007 | COR2-B | 4 | 3 S/FN, 1 FL | 0 |
| May 2007 | COR2-B | 6 | 1 S/FN, 4 FL, 1 OVR | 0 |
| Dec 2007 | COR2-A | 2 | 1 FL, 1 NR/FN | 0 |
| | COR2-B | 7 | 7 NR/ FN | 0 |
| April 2009 | COR2-A | 2 | 1 NR/FN | 1 |
| | COR2-B | 6 | 1 S/FN, 1 NR/FN, 1 FL | 3 |
| | LASCO | 2 | 1 NR/FN | 1 |
| May 2011 | COR2-A | 13 | - | 13[a] |
| | COR2-B | 17 | 3 NR/FN,1 FN | 13[b] |
| | LASCO | 16 | - | 16[c] |

[a,b] seen only in LASCO – (11 NR/FN, 1 FL, 1 FN )
[c] seen only in A&B – (13 NR/FN, 3 FR)
* CME characteristics as seen from the detecting telescope: G: GAP, FL: failed, FN: faint, NR: narrow (<~40°), OVR: overlapping, S: slow (<200 km/s), FR: flux-rope CME

### 3.2. FOV Effects

A comprehensive analysis of VF effects must consider the coronagraph FOV. This is a rarely discussed subject in this context (St. Cyr et al (2000) is the most recent discussion on the matter), yet the FOV largely defines which CMEs will be detected by a given instrument and hence biases the estimates of CME rates and any further studies that rely on CME detectability.

Taking advantage of the availability of multiple coronagraphs operating concurrently, we undertook a careful analysis of FOV effects. For that, we compare CME detections from different coronagraphs on the *same* spacecraft to control for projection effects. In particular, we compare COR1 and COR2 detections for STA and STB, and C2 and C3 detections for SoHO.

We present the results in Table 3. To make the table as concise as possible, we report the events in two groups: events detected in the inner corona ('COR1 or C2') and events detected in the middle/outer corona ('COR2 or C3'). The events detected only in one of the two groups are shown under the 'Only in…' column heading. The effect of the instrument FOV can be assessed by scanning across a row for a given spacecraft ('A' for STA, 'B' for STB, and 'L' for LASCO). For example, we can see that COR1-A detected 18 events in April 2007, COR2-A 16, and 3 (1) events were detected only in COR1-A (COR2-A).

A couple of things are immediately obvious. First, there are no events unique to C3, as found by St. Cyr et al (2000). Either there are no CMEs that manifest themselves in the middle corona (say, by delayed pileup above 6-7 Rs) or C3 lacks the sensitivity to pick certain faint events (for they would have to be faint to be missed by C2) or some events are too slow to leave signatures in synoptic cadence observations (~15-24 mins). Our experience with LASCO data analysis, suggests that all three of these assertions can be true but only for rare types of SBO-CMEs (i.e., so-called 'jaws', see Vourlidas & Webb 2018).

Second, the number of events detected solely in the inner corona increases considerably with solar activity. For May 2011, for example, around 40% of COR1 events are not seen in COR2. The percentages are lower for LASCO. However, COR1 has a lower inner FOV cutoff (1.5 Rs) than LASCO C2 (2.2 Rs), which may be partially responsible for the discrepancy, as we discuss later.

Table 3 CME detections across coronagraphs on the same spacecraft but with different FOVs.

| Date | S/C | Events in Inner FOVs (COR1/C2) | Events in Outer FOVs (COR2/C3) | Only in Inner FOVs (COR1/C2) | Only in Outer FOVs (COR2/C3) |
|---|---|---|---|---|---|
| April 2007 | A | 18 | 16 | 3 | 1 |
| | B | 17 | 12 | 6 | 1 |
| | L | 16 | 13 | 3 | 0 |
| May 2007 | A | 44 | 35 | 10 | 1 |
| | B | 45 | 29 | 16 | 0 |
| | L | 32 | 27 | 5 | 0 |
| Dec 2007 | A | 27 | 20 | 7 | 0 |
| | B | 21 | 12 | 9 | 0 |
| | L | 22 | 17 | 5 | 0 |
| April 2009 | A | 23 | 19 | 7 | 3 |
| | B | 19 | 14 | 8 | 3 |
| | L | 19 | 15 | 4 | 0 |
| May 2011 | A | 119 | 74 | 47 | 2 |
| | B | 121 | 73 | 48 | 0 |
| | L | 75 | 63 | 12 | 0 |

### 3.2.1. COR1-COR2 Detection Discrepancies

As we did in Section 3.1, we delved into the morphology and nature of the events missed by a given telescope in an effort to understand how the FOV may affect CME detectability. In Table 4, we list the characteristics of the missed events, following the order of Table 3 and the letter codes discussed in Sec. 3.1, with one addition. 'OFL' stands for 'outflow'--- an event that lacks the coherent structure of, say, a flux-rope CME and looks like outflowing material (see also Vourlidas et al. 2017).

Six of the 11 events detected only in COR2 are faint, three are outflows, and the remaining two are 'failed' events. The majority of them (7) were detected in COR2-A that has a lower background than COR2-B (Frazin et al. 2012). Faint events are likely missed in COR1 due to the lower SNR compared to the COR2 observations. The slow event in May 2007 was missed because the CME did not appear clearly until well into the COR2-A FOV. This is not unusual for these events as they tend to pileup material high in the corona. The combination of the relatively low sensitivity and a small FOV (4 Rs) is likely to blame. This does not seem to be an issue for the C2 FOV, which extends to 6 Rs, as suggested by the absence of any C3-only events. Of course, the higher C2 SNR may also play a role as well as the small number of examined periods. Overall, we can attribute the grand majority of missed events in COR1 to a combination of instrumental sensitivity and the intrinsic faintness of the events and not the effect of different FOVs between COR1 and COR2.

There is a different story for the events detected in COR1 but not in COR2. Fully 35% (161/454) of them were not detected in the middle corona. The ratio rises to 40% for the high activity period of May 2011. Only 6 of them can be attributed to COR2 data gaps. We assume that those would have been detected in COR2 and we will not consider them further.

Table 4 Characteristics of CMEs missed by telescopes on the same spacecraft. The letter codes are the same as in Table 2. 'OFL' stands for 'outflow'.

| Date | S/C | Events seen only in COR2 or C3 | Total | Events seen only in COR1 or C2 | Total |
|---|---|---|---|---|---|
| April 2007 | A | 1 FN | 1 | 2 OFL, 1 NR | 3 |
|  | B | 1 FN | 1 | 1 OFL, 1 NR, 1 FL. 3 S/FN | 6 |
|  | L | - | 0 | 1 FN, 1 FL, 1 OVR | 3 |
| May 2007 | A | 1 S/FN | 1 | 6 OFL, 1 FL, 2 FN, 1 OVR | 10 |
|  | B | - | 0 | 8 OFL, 3 FL, 2 OVR, 3 S/FN | 16 |
|  | L | - | 0 | 4 FL, 1 OVR | 5 |
| Dec 2007 | A | - | 0 | 5 OFL, 1 NR/FN. 1 GAP | 7 |
|  | B | - | 0 | 4 OFL, 4 NR/FN, 1 GAP | 9 |
|  | L | - | 0 | 3 FN, 2 FL | 5 |
| April 2009 | A | 2 OFL, 1 FL | 3 | 1 OFL, 2 FL, 2 NR, 1 OVR, 1 GAP | 7 |
|  | B | 1 OFL, 2 FN | 3 | 3 OFL, 4 FL, 1 NR | 8 |
|  | L | - | 0 | 1 FL, 3 FN | 4 |
| May 2011 | A | 1 NR/FN, 1 FL | 2 | 11 OFL, 26 FL, 4 NR/FN, 3 OVR, 3 GAP | 47 |
|  | B | - | 0 | 13 OFL, 22 FL, 6 NR/FN, 7 OVR | 48 |

|   | L | - | 0 | 7 FN, 3 FL, 2 OVR | 12 |

The majority of the events missed in COR2 are outflows (54) and 'failed' events (64). The latter, as we discussed earlier and in Vourlidas et al. (2013), are not CMEs and are not expected to survive to the middle corona. The outflows are actually just that; outflows seen in the trailing end of a CME. But why are they counted as individual events in COR1? The answer lies in our approach. To evaluate the VF for an instrument, we must treat its observations *alone* and without the benefit of the broader context provided by the *larger FOV* of another coronagraph. As a CME exits the FOV, and without supporting information from above or below, an observer will have to record each outward-moving signature as an event, particularly if they want to avoid inserting their own judgement in the process. This is, for example, the approach followed for the CDAW catalog. In our case, we have the benefit of observations from an instrument with a large FOV, or more precisely, with a FOV in the middle corona. Hence, we are able to identify the morphological type of the events. Although we followed the same approach as the CDAW catalog in identifying the events in COR1, we do use the additional information to 'clean' the list, as we discussed above. Since, in the end, we do not treat COR1 as an individual telescope, we refrain from deriving a VF for the instrument. The COR2 VF, discussed in Sec. 4, represents both instruments, the same way that the LASCO VF represents C2 and C3. We expand on the implications in Section 4.

Many of the remaining events could have been missed in COR2 due to evolutionary effects. For example, we mark 17 COR1 events (12 of them during high activity) that could not be confidently traced to COR2 due to overlapping CMEs. There are several narrow (in COR1) events that may be missed in COR2 due to the different spatial resolutions and/or due to rapid expansion of these small events, which reduces their brightness. Overall, the vast majority of the COR1-only events can be attributed to non-CMEs and the effects of the relatively narrow FOV that prevents an observer from making a proper assessment of whether an event is an ejection or belongs to a previous event.

### 3.2.2. C2-C3 detections Discrepancies

In the case of LASCO, there are no C3-only events and 29 C2-only events (12 of them in May 2011). There are 11 'failed' events, 4 are missed due to overlapping events and 14 are faint so their lack of C3 signatures could be an effect of the higher instrumental background or of rapid expansion. Recall that the detections are based on standard image processing of synoptic cadences. It is likely that many of the faint C2-only events (and COR1-only) could be detected with more effort. Overall, the effect of the FOV on CME detectability is weaker in LASCO than in SECCHI, given the lower FOV cutoffs of C2 (2.2 Rs, similar to the 2.5 Rs for COR2) and C3 (3.8 Rs). St. Cyr et al. (2000) reported that only ~17% of the CMEs during solar minimum and the rising activity phase (1996-1998) could not be tracked from C2 into C3 field of view. Remarkably, this is the same proportion (29/168 or 17.2%) we find based in our 5-month study that includes events close to Cycle 24 peak.

## 4. Discussion

The compilation of the CME lists and accompanying discussions of our team on the missing events and their morphologies have raised several issues on CME detections, and CMEs in general. We will touch on some of them in this section, starting from the central one: what is the value of VF and how does it affect CME rates?

### 4.1. Visibility Function

| Table 5 Coronagraph Visibility Function for the studied periods assuming a 95% confidence interval | | | | |
|---|---|---|---|---|
| Period | # CMEs | COR2A | COR2B | LASCO |
| Apr 2007 | 16 | 0.80 --1 | 0.50 – 0.90 | 0.80 --1 |
| May 2007 | 35 | 0.90 – 1 | 0.67 – 0.92 | 0.90 – 1 |
| Dec 2007 | 22 | 0.72 – 0.97 | 0.47 – 0.84 | 0.85 – 1 |
| Apr 2009 | 22 | 0.72 – 0.97 | 0.51 – 0.87 | 0.72 – 0.97 |
| **May 2011** | **91** | **0.77 -- 0.92** | **0.71 – 0.87** | **0.73 – 0.89** |

The VFs in Table 1 are based on the underlying assumption that the unique events in each month are also the total number of CMEs ejected during that period. In other words, we assume that no CME was missed by all coronagraphs. Obviously, this assumption needs justification, given, particularly, the role projection effects play in CME detection. For example, it is a weak assumption for 2007, when the three spacecraft are at relatively small angular separations from each other and thus have similar viewpoints to the Sun. It becomes uncertain during periods of high activity, when the overlap of events becomes more likely. Intermediate situations with moderate solar activity and quadrature viewpoint configurations may be the best periods for VF evaluation. April 2009 offers the quadrature advantage but solar activity was still at minimum. We note again, that we are referring to CME detections based on standard analysis, using running differences (or similar schemes) on synoptic images. In our experience, even very faint CMEs (Nieves-Chinchilla et al. 2013) and complex overlapping situations (Colaninno & Vourlidas 2015) can be identified with sufficient investment of effort. But these are specialized cases and are not really amenable to catalog building from which VFs are derived.

It seems rather optimistic to expect that a period with optimal spacecraft and event separation exists and, in any case, the VF is expected to vary as the instrument ages (e.g. MacQueen et al. 2001) and may be somewhat dependent on the level of activity, if event overlap becomes important. So, we turn to a statistical approach for capturing the uncertainty in the total number of CMEs for a given period. Since we are dealing with small samples and a binomial problem ('detection' vs. 'non-detection'), we adopt a Bayesian approach that is optimal for such problems, as detailed in Cameron (2011). We use a 95% Confidence Interval (CI) that results in the VFs shown in Table 5. Our initial findings (3.1) remain unchanged—COR2A and LASCO have similar VFs, COR2B has a lower VF. We see a decrease in the VF for May 2011 due to the high activity, as we discussed in the previous paragraph. It corresponds, however, to a period of wide longitudinal coverage of the Sun with the coronagraphs in quadrature and thus the number of unique CMEs may be closer to the true number of events. Hence, the VFs for May 2011 are likely more representative of the actual instrument VFs compared to the other periods. It may be no accident that all three coronagraphs have similar VFs during this period.

## 4.2. Corrected CME Rates

The main application for a VF is the correction of CME rates to allow studies of the correlation of CMEs to other types of solar activity, including the variation of mass flux into the heliosphere. So, we proceed to apply the VFs from Table 5 to the observed uncorrected CME rates in LASCO and SECCHI/COR2 and examine the effect. An additional important correction is the duty cycle of the instrument, defined as the ratio of the instrument observing time over a time period of interest (e.g. daily, Carrington rotation) and is an essential correction for observations from the ground or low-Earth orbits (e.g. Webb & Howard, 1994; St. Cyr et al. 2015). We have calculated previously the duty cycles for both LASCO (Vourlidas et al. 2010) and COR2 (Vourlidas et al. 2017). Table 6 summarizes the corrections and final rates for each of the five periods we analyzed.

Table 6 Daily CME rates corrected for the coronagraph duty cycle (form Vourlidas et al. 2017) and visibility function. The 95% CI concerns the VFs only (Table 5).

| Coronagraph | Correction | CME Rate | | | | |
|---|---|---|---|---|---|---|
| | | Apr 2007 | May 2007 | Dec 2007 | Apr 2009 | May 2011 |
| COR2A | Observed | 0.53 | 1.13 | 0.65 | 0.67 | 2.52 |
| | Duty Cycle | 0.59 | 1.21 | 0.65 | 0.72 | 2.57 |
| | **VF (95% CI)** | **0.59 – 0.74** | **1.21 – 1.34** | **0.67 – 0.91** | **0.74 – 1.01** | **2.57 – 3.35** |
| COR2B | Observed | 0.40 | 0.94 | 0.48 | 0.53 | 2.35 |
| | Duty Cycle | 0.41 | 0.96 | 0.56 | 0.56 | 2.38 |
| | **VF (95% CI)** | **0.46 – 0.83** | **1.04 – 1.43** | **0.67 – 1.19** | **0.64 – 1.09** | **2.73 – 3.37** |
| LASCO | Observed | 0.53 | 1.13 | 0.71 | 0.67 | 2.42 |
| | Duty Cycle | 0.53 | 1.38 | 0.71 | 0.67 | 2.42 |
| | **VF (95% CI)** | **0.53 – 0.67** | **1.38 – 1.53** | **0.71 – 0.84** | **0.69 – 0.93** | **2.72 – 3.31** |

We can see that the observed uncorrected rates lie close (~10%) to the lower CI limit of the corrected rates for most cases, except for LASCO on May 2007 (22%), and COR2B on Dec 2007 (~40%) and April 2009 (20%). In Figure 2, we compare the VF-corrected rates to the (already 'duty cycle'-corrected) CME rates from various CME catalogues in Vourlidas et al. (2017). The largest effects are seen in May 2007 and May 2011. Both months were periods of increased activity so the large corrections are expected.

In other words, we should *treat the observed CME rates as lower limits to the actual rates*. On the other hand, the CI upper limit suggests that the global rates may be underestimated by 30-40%. Therefore, we find it is important to consider the effects of both instrument duty cycle and visibility function in studies of solar activity, involving CMEs, as the previous studies have shown (Webb & Howard 1994; St Cyr et 2000, 2015; Webb et al. 2017).

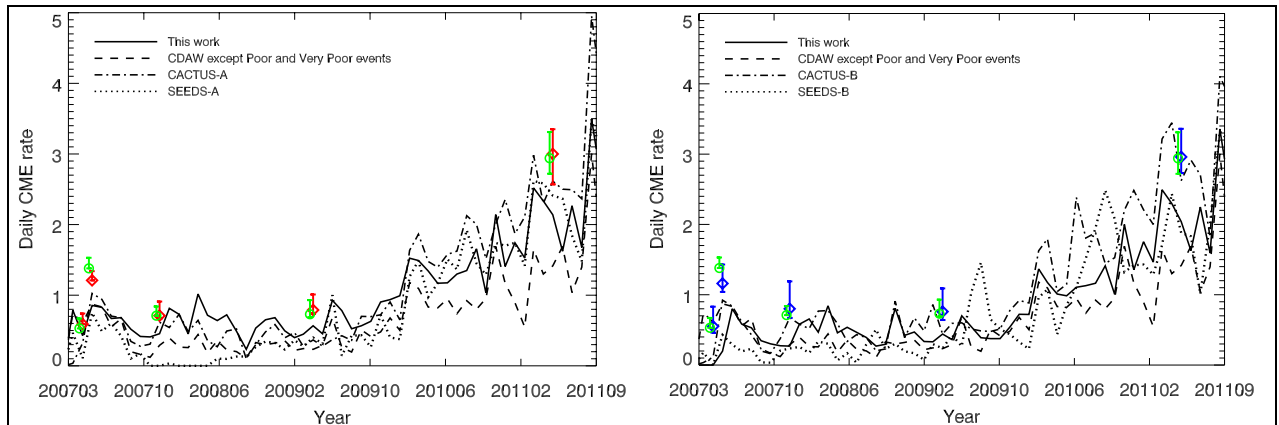

Figure 2 Comparison of the corrected (for DC and VF) CME rates from this work to rates from other catalogues corrected for DC only (adopted from Vourlidas et al. 2017). The error bars correspond to 95% CI. All rates were corrected for duty cycle and averaged over a month

### 4.3. Do missing CMEs matter?

Ideally, we need to know the absolute number of CMEs ejected by the Sun to measure the true VF of a given coronagraph and consequently to derive accurate CME rates and their contribution to the solar mass loss flux. Nominally, this should be straightforward to measure if observations from widely distributed (and equally capable) coronagraphs are available. In lieu of such capability, it is more practical to ask which CMEs are missed from a single viewpoint and whether they matter for the accuracy of the VF and CME rates. Here, the definition of a CME becomes important and our definition in Sec 2.2 drives our assessments below.

As we discussed in Section 3, there are four main ways a CME could be missed: (1) due to the combination of event size and propagation direction, i.e. event narrow enough to remain behind the occulter until it becomes too faint; (2) due to activity, i.e. event overlap, high levels of particle hits; (3) due to telescope sensitivity, i.e. event too faint; (4) due to the observing program, i.e. event too slow. These four factors affect unevenly the VF.

Obviously, (1) requires quite a unique combination of event properties and Sun-observer configuration and must be quite rare. For example, for a CME to be missed by both LASCO coronagraphs, it must remain behind the C3 occulter up to at least 30 Rs, it must be less than 15º wide and be directed right along the Sun-Earth line. For COR2, with occultation at about 3 Rs, the CME would have to be narrower than 23º at a distance of 15 Rs. Of course, observer-directed CMEs stay in the instrument FOV for much larger distances than the actual FOV (30/15 Rs for C3/COR2, respectively) but they will be even fainter, so the estimates we quote above are rough upper limits. Narrow events (based on projected widths) generally show little internal structure, are difficult to detect above 10 Rs or so and appear mostly as outflows behind larger CME events. For these reasons, we do not consider them in our statistics (Vourlidas et al. 2010, 2013, 2017). Now, events with true (not projected) widths below 20º would be even more difficult to classify as CMEs. In any case, even if they were bona-fide CMEs, the special circumstances required for not detecting them argue against narrow events being an important factor in VF uncertainty. Concurrent observations in EUV in the low corona add confidence to the identification of these events directed toward the observer.

Event overlap seems to be of concern only during solar maximum. Based on our experience, the effect is largely mitigated with radial coverage, e.g. LASCO C2+C3 or SECCHI COR1+COR2. Inner corona observations tend to capture the individual events before they merge (in projection) at the outer coronagraphs FOVs. Our investigation (Table 4) suggests that only a handful of events (2-3) are missed in a month during solar maximum, at least in Cycle 24. Cycle 23 was more active with a greater chance of event overlap. A simple way to estimate a missing CME number due to overlap for cycle 23 is to scale with the uncorrected CME rates between 2002 (cycle 23 max; 4.5 CMEs/day) and may 2011 (2.52 CMEs/day). This exercise suggests that up to 5 events may be missed monthly due to overlap, which is not significant enough to affect CME rate estimates or large-scale studies on the effects of CMEs in the heliosphere However, it may be an issue for space weather research where the focus may be on the evolution of individual events and interactions among CMEs.

Clearly, a coronagraph with low SNR is going to miss many events. The higher background of COR2-B compared to COR2-A is a good example of this (Table 1 and Sec. 3.1, see also Vourlidas et al. 2017). Yet, the number of missed events by COR2-B *only* is just a handful (4-7, Table 2). Almost all of them are narrow (hence unlikely to be CMEs, see discussion on item (1) above) and only a quarter of them (5 out of 21) are likely CMEs. Across the telescopes, the only events missed are faint narrow ejections. We argue, based on the limited analysis here, that no real CMEs (as we define in 2.2) were missed by the modern coronagraphs on SOHO and STEREO and hence we expect the same performance for future instruments, as well.

Finally, the concept of operations is going to influence the CME detectability for a particular coronagraph. Wang & Colaninno (2014) showed how the increase in the LASCO cadence since 2010 led to a large increase in the entries in both automated and manual catalogs. Most of these entries are actually post-CME flows and not real CME events. In principle, a higher cadence synoptic program should help resolve event overlap and capture faster events. This is not an issue for the coronagraphs here. From the tables, one can see that slow/faint events are the ones missed. This is a known issue in coronagraph observations; slow, halo-like CMEs are the "Achilles heel" of automated algorithms and are even difficult to identify visually. Generally, they can be detected in very low cadence (~hours) running difference images but such analyses are non-standard and one would have to suspect, in advance, that such an event may be present in order to use such approaches. So, we have to assume that such halo (full or partial) CMEs will be missed given the 'standard' synoptic programs of the LASCO and SECCHI coronagraphs, with cadences of the order of 12-24 minutes. Since such CMEs are much more likely during solar minima when the overall CME rates are low, this could be a significant effect for instrument VF and CME rates. However, our analysis shows only 5 such events in the 5 periods, so it seems to be no more important than event overlap.

Overall, we conclude that modern coronagraphs do not miss any important CMEs (flux-rope CMEs) thanks to their adequate sensitivity and observing cadence. Possible exceptions are slow, out-of-plane (halo-like) CMEs during minima. But given the low occurrence of such events, the effect is minimal on the VF and/or CME rates.

### 4.3.1. Towards Robust CME Rate Measurements

The previous discussion makes the case that coronagraph observations from a single viewpoint are unlikely to capture all CMEs thus injecting uncertainty on the VF and CME rates. Even if all non-CME features (e.g., jets, wave-like ejections) are ignored, there are bound to be operational gaps and event overlap that reduce the duty cycle and the VF of a telescope. How can we ensure, therefore, we capture all CMEs reliably?

The obvious answer is to observe from viewpoints distributed in longitude (or latitude---the effects are similar for the small solar elongations considered here so we will not discuss latitude separately, except when necessary). The next considerations are how many and how far apart.

In the current work, we explored angular separations up to 90º between coronagraphs from the 3 available platforms. The quadrature configuration is important as it was the only case of detection of FR-CMEs. Three FR-CMEs, missed by LASCO due to projection effects, were detected by COR2A/B (Table 2). On the other hand, the two COR2s were at opposition during May 2011, and having essentially the same lines of sight, were effectively reduced to a single viewpoint (Thompson et al. 2011). It is no surprise that both missed the same events (13), although none of those was a major CME, in that month, at least. Therefore, quadrature (90º) is the optimal configuration for CME detection, if only two viewpoints are available. The possibility remains that some events may be missed, either due to overlap during activity or if they are narrow and propagate at large angles from the sky planes of both instruments. In the quadrature case, the latter scenario requires CMEs of about 20º - 30º width propagating at 45º or so from a given sky plane.

In the case of three viewpoints, then the optimal configuration would be equal angular separation of 120º among them, which provides a full coverage of the corona. In this case, it is hard to imagine an event size and/or propagation direction that would put the CME at a significant distance from the sky plane of all three viewpoints, and thus making it difficult to detect. The chance of event overlap is also diminished, enabling separation of both slow events and events in quick succession, as Colaninno & Vourlidas (2015) and Patsourakos et al. (2016) demonstrated with SECCHI/LASCO configurations at 109º-118º apart. These are very important capabilities for Space Weather research. However, disentangling multiple events will almost certainly require more extensive analysis than a simple inspection of movies from the different viewpoints.

We conclude that coronagraphic observations from three viewpoints equally distributed around the Sun are sufficient for detecting the vast majority of CMEs and will provide robust VFs and CME rates, for all practical purposes. Unfortunately, orbital mechanics do not provide stable orbits at 120º from Earth making it challenging to build such an observing system without the use of advanced station keeping technologies (e.g. solar sails). Our analysis suggests that two coronagraphs at quadrature would provide a sufficiently robust CME detection configuration. So, we adopt our VFs results for May 2011 as the formal VF for the COR2 and LASCO instruments. Rather than one viewpoint being along the Sun-Earth line, as it is currently the case, it would be more effective, from a Space Weather perspective, if the two instruments were located at +/- 45º from Earth. In that case, none of the Earth-directed CMEs would appear as halos and their kinematics and other properties could be measured more reliably. However, this configuration is equally difficult to maintain without a complex station keeping solution. Observatories at the $L_4/L_5$ Sun-Earth Lagrange points offer an acceptable and easily implementable alternative.

### 4.4. Instrument considerations

The twin STEREO coronagraphs afford us the unique opportunity to compare coronagraphs of the same design, operating with the same observing strategy and data pipeline against the same CME sample. The COR1 inter-calibration (Thompson et al 2008; 2011) demonstrated that COR1-A and B have very similar background levels and responses to CMEs, so we did not expect to see differences in COR1 CME detectability, which Table 3 confirms.

This is not true for COR2, though. COR2-B consistently misses more CMEs than COR2-A (Tables 2 and 3) and thus results in lower (uncorrected) CME rates (Table 6). The issue arises from the higher background stray light level in COR2-B and has been discussed extensively in the context of other coronagraphs (Frazin et al. 2012) and CME detection (Vourlidas et al. 2017). However, we note that the application of the VF derived in this paper brings the COR2-B CME rate in agreement (with somewhat wider CI) with the CME rates from COR2-A and LASCO (Table 6) and thus provides further confidence in our methodology.
We conclude that the COR1 and COR2 A/B pairs perform very similarly, despite the albeit small differences in their design (i.e. different occulter sizes to account for the 10% difference in the heliocentric distance between STEREO-A and B) and that accounting properly for stray light levels is all that is needed to cross-calibrate their detection rates.

### 4.5. The role of FOV

The cross-check of CME detections from instruments along the same vantage point but with different FOVs resulted in one key finding: the FOV does matter. In particular, a low inner FOV cutoff can help discriminate between different events in periods of high activity but also tends to inflate "CME" rates by detecting lots of apparent ejections that may not be CMEs. A wide FOV helps detect slow and faint CMEs, particularly halos. These findings confirm the suggestions by St Cyr et al (2000) based on the early LASCO CME detection and have important Space Weather ramifications.

FOVs that extend to 10-12 Rs and beyond are crucial for establishing whether an event escapes into the heliosphere or not and hence whether it is a CME (Sec 2.2 and also Vourlidas et al. 2013). Naturally, the design of a coronagraph is optimized to the particular objectives of the mission. If CME detection is the driver, as is for some Space Weather research then the FOV size and inner cutoff need to be carefully considered.

Typically, space-based coronagraphs for the inner corona cover from 1.5 to 3-4 Rs (e.g. COR1, C1, SMM; the ground-based MLSO coronagraphs have an even lower inner field-of-view, but at the expense of reduced altitude coverage. This is the region where the initial acceleration of the CME takes place, which is important for Solar Energetic Particle (SEP) alerts (St. Cyr et al. 2017). But these FOVs are too small to reliably detect all CMEs, especially slow SBO-CMEs and halo CMEs. Our VF results in Table 1 and the findings in Sec. 3 strongly suggest that a coronagraph with inner FOV cutoff about 2 Rs and, at least 10 Rs width would detect almost all CMEs from that viewpoint and will easily discriminate between MFR-CMEs and other outflows or waves. COR2 comes close to this ideal design as does the Compact Coronagraph (CCOR; Gong & Socker 2004) to be deployed by NOAA in 2024 as the first operational coronagraph. If the CME acceleration and/or the generation of SEPs are design drivers, the inner FOV cutoff should be reduced to 1.5 Rs or even lower, if possible.

The FOV findings have implications for the existing CME catalogs, which are designed to record every outgoing density/brightness disturbance in the solar wind flow without consideration on whether it is a CME, part of a CME, or just a wave. This approach is useful for compiling solar wind transient event lists but not for compiling CME lists. In our discussion of Table 3, we pointed out that 40% of the COR1 detections were nor related to individual CMEs. Indiscriminately accepting those events as CMEs leads to incorrect analyses. For example, CME rates tend to be inflated while average properties of CMEs tend to be biased. In an earlier analysis, we found that one third of the entries in the CDAW catalogue are incorrect (Vourlidas et al. 2013). This is the reason that we chose to not use the catalog and identified every event ourselves. This is also the reason we do not provide a COR1 VF (sec 3.2.1).
In conclusion, we find that the FOV of a coronagraph plays a critical role in properly identifying CMEs. Small FOV coronagraphs have unreliable VFs.

## 5. Conclusions

We have analyzed simultaneous CME observations from three viewpoints over 4 monthly periods. The periods were selected because of unique orbital configurations that allow us to investigate projection and instrument performance effects on the ability of a coronagraph to detect CMEs and thus provide reliable Visibility Functions and CME rates. The four periods corresponded to (1) April-May 2007 when all three telescopes had very similar LOS; (2) Dec. 2007 when SECCHI-A and B were 40º apart (project effects may be important) but each SECCHI and LASCO were only 20º apart (minimal projection effects); (3) April 2009 when SECCHI-A and B were in quadrature; (4) May 2011 when SECCHI-A and B are in opposition but they are also in quadrature with LASCO. For each configuration, we visually examined all available observations in great detail to construct event catalogs, to morphologically classify each event, and to identify the number of unique events for each period. We then proceeded to derive the VF for each telescope (COR1/COR2-A and B, C2, C3) and investigate the nature and reasons behind the missed events in each case.

Our main findings are summarized below:

- We derive the following VFs with 95% Confidence Interval: C2+C3 (0.73- 0.89), COR2A (0.77 – 0.92), COR2B (0.71 – 0.87).
- The VFs are instrument- and activity-dependent (Sec. 4) and may vary in time. We find that high levels of solar activity may lower the VFs but the missing events are likely small and less energetic events. However, the strength of these findings can be assessed reliably only through an extension of the current approach to a larger period covering a significant part of the cycle, say, at least 4 years.
- CME detection varies significantly among instruments due to optical design and field-of-view difference that lead to differences in instrumental stray light. Hence, every coronagraph is unique, even when they have the same design (i.e. COR1A/B, COR2A/B).

- The FOV of a coronagraph is critical for CME detection. Small FOV coronagraphs have uncertain VFs and CME rates. For this reason, we do not derive VFs for COR1 alone here.
- COR2A/B (FOV <15 Rs) and LASCO (FOV <30 Rs) have very similar VFs (Table 5). Hence a FOV of about 15 Rs is sufficient for CME detection.
- After examining the morphology and kinematic characteristics of all missed CMEs (Table 2 & Table 4), we conclude that almost none of the missed events were real CMEs (in the sense they had flux-rope structure, following the Vourlidas et al. (2013) definition). The only CME candidates were slow events (S/FN in Tables 2 and 4), during solar minimum, that were likely streamer-blowout CMEs. Those can be detected with low cadence running difference movies (e.g. at 1-2 hour cadence). We suggest that such movies should be inspected (and be made into standard data products) when high detection accuracy is required. Narrow events (<20º) are the most common type for missed CME. The remaining features (outflows, waves) are not real CMEs so they are of no concern here.
- The rather obvious, but worth repeating, finding is that the detectability of a CME is governed by its brightness *as viewed from a given vantage point*. Hence, Thomson scattering effects need to be considered when deploying instrumentation. For example, if the detection of Earth-directed CMEs is a key objective for an instrument, Sun-Earth line deployment is not optimal.
- Two viewpoints in quadrature will miss very few CMEs (mostly because of event overlap during solar maximum) and thus can provide reliable measurements of CMEs rates. The optimal location (for Space Weather considerations) would be +/- 45º from the Sun-Earth line. However, the more stable L4/L5 locations are acceptable and easier to design for.
- Correction for VF increases CME rates by up to 40%, although a 10% is more common across the cycle. The uncorrected CME rates should be considered *as lower limits*.

## Acknowledgements

We thank the referee for useful comments that have improved the manuscript. The work was primarily funded by NASA HGI grant NNX17AC47G. The SECCHI data are produced by an international consortium of the NRL, LMSAL and NASA GSFC (USA), RAL and University of Birmingham (UK), MPS (Germany), CSL (Belgium), and IOTA and IAS (France). SOHO is a project of international cooperation between ESA and NASA.